\documentclass[conference]{IEEEtran}
\IEEEoverridecommandlockouts
\usepackage{cite}
\usepackage{amsmath,amssymb,amsfonts}
\usepackage{algorithmic}
\usepackage{graphicx}
\usepackage{textcomp}
\usepackage{xcolor}
\def\BibTeX{{\rm B\kern-.05em{\sc i\kern-.025em b}\kern-.08em
    T\kern-.1667em\lower.7ex\hbox{E}\kern-.125emX}}
\begin{document}

\title{Accurate Congenital Heart Disease Model Generation for 3D Printing  
}

\author{

\IEEEauthorblockN{Xiaowei Xu, Tianchen Wang, Dewen Zeng, Yiyu Shi}
\IEEEauthorblockA{\textit{Department of Computer Science and Engineering} \\
\textit{University of Notre Dame}\\
South Bend, IN, USA \\
\{xxu8, twang9, yshi4\}@nd.edu}
\and
\IEEEauthorblockN{Qianjun Jia, Haiyun Yuan, Meiping Huang, Jian Zhuang}
\IEEEauthorblockA{
\textit{Cardiovascular Surgery Department}\\
\textit{Guangdong General Hospital}\\
Guangzhou, China \\
meipinghuang.gdg@gmail.com}

}

\maketitle

\begin{abstract}
3D printing has been widely adopted for clinical decision making and interventional planning of Congenital heart disease (CHD), while whole heart and great vessel segmentation is the most significant but time-consuming step in model generation for 3D printing.
While various automatic whole heart and great vessel segmentation frameworks have been developed in the literature, they are ineffective when applied to medical images in CHD, which have significant variations in heart structure and great vessel connections. To address the challenge, we leverage the power of deep learning in processing regular structures and that of graph algorithms in dealing with large variations, and propose a framework that combines both for whole heart and great vessel segmentation in CHD.
Particularly, we first use deep learning to segment the four chambers and myocardium followed by blood pool, where variations are usually small. We then extract the connection information and apply graph matching to determine the categories of all the vessels.
Experimental results using 68 3D CT images covering 14 types of CHD show that our method can increase Dice score by 11.9\% on average compared with the state-of-the-art whole heart and great vessel segmentation method in normal anatomy.  
{The segmentation results are also printed out using 3D printers for validation. }
\end{abstract}


\section{Introduction}

Congenital heart disease (CHD) is the problem with the heart’s structure that is present at birth, which is the most common cause of infant death due to birth defects \cite{bhat2016illustrated}.
It usually comes with significant variations in heart structures and great vessel connections, which makes it time-consuming, tedious, and low-reproductivity to manually process (e.g., segment, diagnose, analyzed) 3D medical images.
Recently, three-dimensional (3D) printing has been widely adopted in clinic, which is useful in clinical decision making, interventional planning, facilitating communication between physicians and patients, as well as enhancing medical education for a variety of learners.
However, the main step of model generation for 3D printing, whole heart and great vessels segmentation is rather time-consuming and labor-intensive which takes an experienced radiologist hours to produce only one 3D CHD segmentation.
Thus, considering the large quantity of medical images and the increasing cost of medical expense \cite{xu2018resource}\cite{xu2018quantization}, automatic whole heart and great vessel segmentation of heart in CHD is emerging.

Recently, the development of deep learning \cite{xu2018scaling}\cite{tianchen2019}\cite{xu2018efficientHardware} has improve the performance of segmentation \cite{xu2017edge}\cite{li2019exploiting}\cite{liu2019compression} including whole hearts and great vessels segmentation by a large margin. 
One approach is about multi-modality whole heart segmentation \cite{zhuang2016multi} which deals with seven substructures within normal heart anatomy. 
There are tens of works \cite{wang2018two}\cite{payer2017multi}\cite{xu2018cfun} in this approach, 
and the state-of-the-art performance is obtained by \cite{payer2017multi} which combines 3D U-net \cite{cciccek20163d} for segmentation and a simple convolutional neural network for label position prediction.
Another approach is about blood pool segmentation in CHD which only handles the blood pool and myocardium \cite{yu20163d}\cite{wolterink2016dilated}.
There are also some works leveraging user interaction for accurate segmentation. For example, Danielle \textit{et al.} \cite{pace2018iterative} adopted iterative segmentation for left ventricle (LV) and aorta segmentation in CHD, which required user interaction to locate an initial seed for segmentation.
Considering the significant variations in heart structures and great vessel connections in CHD, almost all the existing methods cannot effectively perform whole heart and great vessels segmentation in CHD.
Inspired by the success of graph matching in a number of applications with large variations \cite{lajevardi2013retina}, in this paper we propose to combine deep learning and graph matching for fully automated whole heart and great vessel segmentation in CHD.
Particularly, we leverage deep learning to segment the four chambers and myocardium followed by blood pool, where variations are usually small and accuracy can be high. We then extract the vessel connection information and apply graph matching to determine the categories of all the vessels. We collected 68 CT images with 14 types of CHD for experiment.
Compared with the state-of-the-art method for whole heart and great vessel segmentation in normal anatomy, our method can achieve 11.9\% higher Dice score. 
{We further print out the segmentation results on 3D printers for validation.}





\begin{figure*}[!tb]
\includegraphics[width=\textwidth]{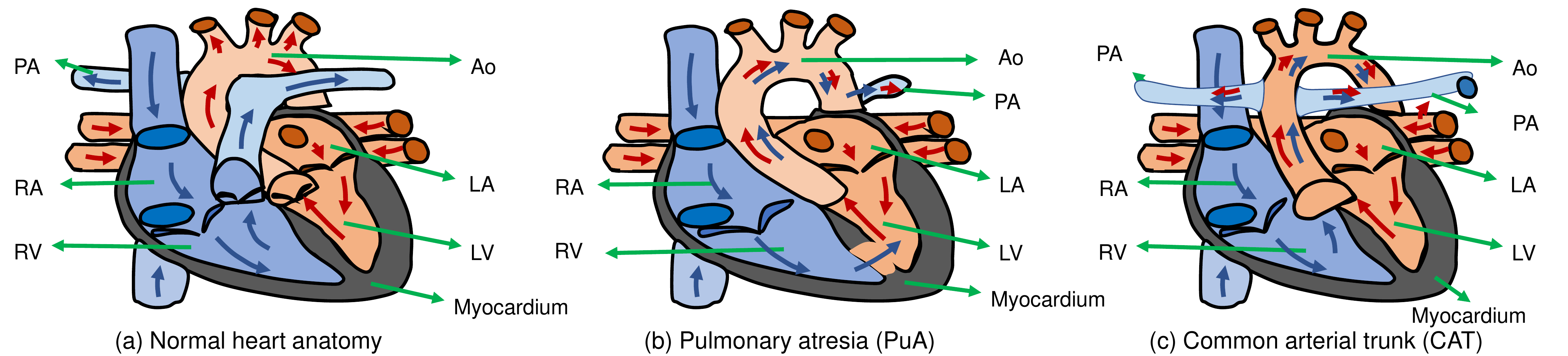}
\vspace{-12pt}
\caption{Examples of large structure variations in CHD. In normal heart anatomy (a), PA is connected to RV. However, in pulmonary atresia (b), PA is rather small and connected to descending Ao. In common arterial trunk (c), Ao is connected to both RV and LV, and PA is connected to Ao.
} 
\label{fig1}
\end{figure*}
\section{Background}\label{sec_background}

\begin{figure*}[!tb]
\includegraphics[width=\textwidth]{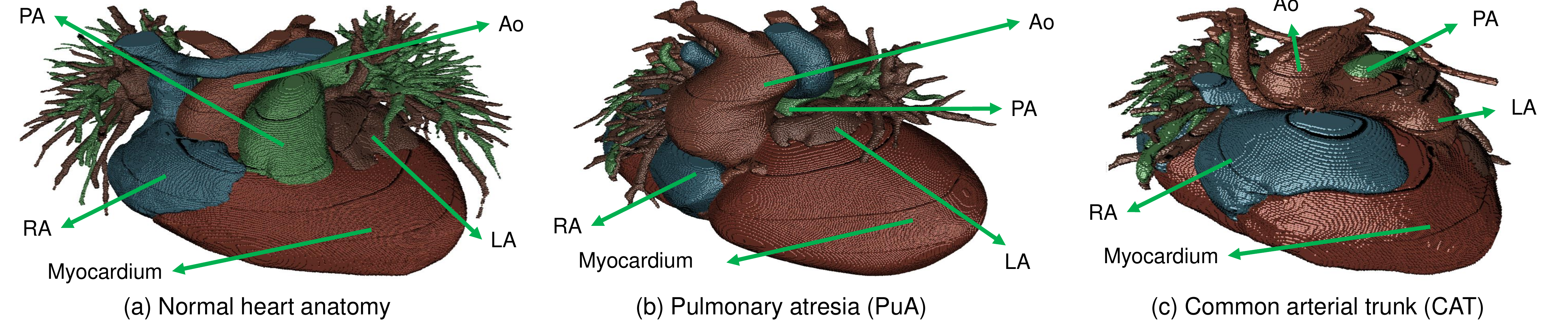}
\vspace{-12pt}
\caption{Pulmonary atresia and common arterial trunk examples in our dataset, with large variations from normal heart anatomy.
} 
\vspace{-12pt}
\label{fig2_dataset}
\end{figure*}

\begin{table*}
\centering
\caption{The types of CHD in our dataset and the associated number of images. Note that some images may correspond to more than one type of CHD.}
\begin{tabular}{cccccccc|cccccc}
\hline
\multicolumn{8}{c}{Common CHD} & \multicolumn{6}{c}{Less Common CHD}\\\hline
ASD&AVSD&VSD&FoV&PDA&TGA&CA&PS&PAS&AD&CAT&AAA&SV &PuA  \\
17&4&26&7&7&4&4&4&3&20&4&8&2&7\\
\hline
\end{tabular}
\vspace{-12pt}
\label{dataset}
\end{table*}

Within normal heart anatomy as shown in Fig.~\ref{fig1}(a), there are usually seven substructures: left ventricle (LV), right ventricle (RV), left atrium (LA), right atrium (RA), myocardium (Myo), faorta (Ao) and pulmonary artery (PA).
Note that the area including RA, LA, LV, RV, PA, and Ao is defined as blood pool.
However, CHD usually suffers from significant variations in heart structure and great vessel connections.
Eight common types of CHD \cite{bhat2016illustrated} include: atrial septal defect (ASD), atrio-ventricular septal defect (AVSD), patent ductus arteriosus (PDA), pulmonary stenosis (PS), ventricular septal defect (VSD), co-arctation (CA), Tetrology of Fallot (ToF), and transposition of great arteries (TGA).
Fig.~\ref{fig1}(b)(c) shows two less common types with larger variations, where we can notice that PA is connected to Ao rather than RV. 
As existing methods perform pixel-wise classification based on the surrounding pixels in the receptive field, the disappeared main trunk of PA renders them ineffective.



\section{Dataset}
Our dataset consists of 68 3D CT images captured by a Simens biograph 64 machine.  The ages of the associated patients range from 1 month to 21 years, with majority between 1 month and 2 years. The size of the images is $512\times 512\times$(130$-$340), and the typical voxel size is 0.25$\times$0.25$\times$0.5$mm^3$. The dataset covers 
14 types of CHD (out of 16 total \cite{bhat2016illustrated}), which include the eight common types discussed in Section \ref{sec_background}
plus six less common ones (pulmonary artery sling (PAS), anomalous drainage (AD), common arterial trunk (CAT), aortic arch anomalies (AAA), single ventricle (SV), pulmonary atresia (PuA)). The number of images associated with each is summarized in Table \ref{dataset}. All labeling were performed by experienced radiologists, and the time for labeling each image is 1-1.5 hours.
The labels include seven substructures: LV, RV, LA, RA, Myo, Ao and PA. 
For easy processing, venae cavae (VC) and pulmonary vein (PV) are also labeled as part of RA and LA respectively, as they are connected and their boundaries are relatively hard to define. Anomalous vessels are also labeled as one of the above seven substructures based on their connections. Fig. \ref{fig2_dataset} shows 3D views of some examples in our dataset with significant structure variations.

\begin{figure*}[!tb]
\includegraphics[width=\textwidth]{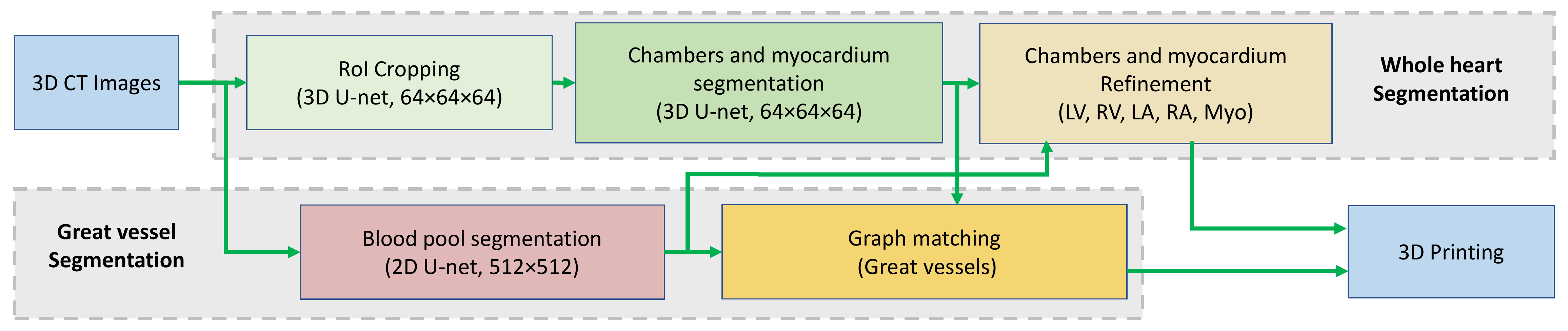}
\vspace{-20pt}
\caption{Overview of the proposed framework combining deep learning and graph matching for whole heart and great vessel segmentation in CHD.
} 
\label{fig_framework}
\end{figure*}

\begin{table*}
\centering
\caption{Mean and standard deviation of Dice score of the state-of-the-art method Seg-CNN \cite{payer2017multi} and our method (in \%) for 14 types of CHD in whole heart and great vessel segmentation.}
\begin{tabular}{c|cccccccc|cccccc}
\hline
Method&\multicolumn{8}{c}{Common CHD} & \multicolumn{6}{c}{Less Common CHD}\\
&ASD&AVSD&VSD&FoV&PDA
&TGA&CA&PS&PAS&AD
&CAT&AAA&SV &PuA  \\\hline
Seg-CNN \cite{payer2017multi} 
&66.4&70.1&67.3  &68.7 &71.4
&61.0  &66.1   &65.9   &65.9   &66.7
&61.0  &66.3  &62.6  &66.3 \\
with Std&$\pm$4.6  &$\pm$3.2  &$\pm$4.7  &$\pm$1.4  &$\pm$2.6
&$\pm$13.3 &$\pm$3.3  &$\pm$4.5  &$\pm$9.9  &$\pm$5.0
&$\pm$4.1  &$\pm$5.0  &$\pm$8.4  &$\pm$6.0\\\hline
Our method
&78.5  &83.1  &77.6   &82.1   &82.8 
&77.1&79.2 &75.9 &78.9 &78.4
&75.6  &77.8  &74.3&71.8\\
with Std &$\pm$4.1  & $\pm$4.5  &$\pm$8.5  &$\pm$2.4  &$\pm$3.8
&$\pm$7.0  &$\pm$4.2  &$\pm$4.0  &$\pm$3.8  &$\pm$5.4
&$\pm$8.8  &$\pm$5.1  &$\pm$3.3  &$\pm$11.8\\ 
\hline
\end{tabular}
\vspace{-12pt}
\label{comparison}
\end{table*}

\section{Method}

\subsection{Framework Overview}
The overall framework is shown in Fig. \ref{fig_framework}.
\textbf{Region of interest (RoI) cropping} extracts the area that includes the heart and its surrounding vessels. We resize the input image to a low resolution of 64$\times$64$\times$64, and then adopt the same segmentation-based extraction as \cite{payer2017multi} to get the RoI. 
\textbf{{Chambers and myocardium segmentation}} resizes the extracted RoI to 64$\times$64$\times$64 which is fed to a 3D U-net for segmentation.
\textbf{{Blood pool segmentation}} is conducted on each 2D slice of the input using a 2D U-net with an input size of 512$\times$512. 
Note that in order to detect the blood pool boundary for easy graph extraction in graph matching later, we add another class blood pool boundary in the segmentation.
\textbf{{Chambers and myocardium refinement}} refines the boundaries of chambers and myocardium based on the outputs of chambers and myocardium segmentation and blood pool segmentation.
\textbf{{Graph matching}} identifies great vessels and anomalous vessels using the outputs of blood pool segmentation and chambers and myocardium segmentation.

\section{Experiments}

\subsection{Experiment Setup}
All the experiments run on a Nvidia GTX 1080Ti GPU with 11GB memory.
We implement our 3D U-net using Pytorch based on \cite{payer2017multi}.
For 2D U-net, most configurations remain the same with those of the 3D U-net except that 2D U-net adopts 5 levels and the number of filters in the initial level is 16.
Both Dice loss and cross entropy loss are used, and the training epochs are 6 and 480 for 2D U-net and 3D U-net, respectively.
Data augmentation is also adopted with the same configuration as in \cite{payer2017multi} for 3D U-net.
Data normalization is the same as \cite{payer2017multi}.
The learning rate is 0.0002 for the first 50\% epochs, and then 0.00002 afterward. We adopt Seg-CNN \cite{payer2017multi} that achieves the state-of-the-art performance in normal anatomy whole heart and great vessel segmentation for comparison. 
The configuration is the same as that in \cite{payer2017multi}. 

For both methods, four-fold cross validation is performed (17 images for testing and 51 images for training). The split of our dataset considers the structures of CHD so that any structure in the testing dataset also has a similar presence in the training dataset, though they may be not of the same type of CHD. The Dice score is used for segmentation evaluation.

{We finally print out part of the segmentation results on a commercial 3D printer Sailner J501Pro for validation. 
It usually takes 3-4 hours to print a model (segmentation result) of children’ hearts.}


\subsection{Results and Analysis}
The comparison with Seg-CNN \cite{payer2017multi} is shown in Table \ref{comparison}.
Our method can get 5.5\%-16.1\% higher mean Dice score across the 14 types of CHD. (11.9\% higher on average). 
The highest improvement is achieved in TGA, which is due to the fact that both Ao and PA are with normal structures.
The least improvement is obtained in PuA, which is due to the fact that PuA is with serious variation on the structure and connection of PA.
Both Seg-CNN and our method obtain a slightly higher accuracy on common CHD than less common CHD.
Our method achieves a similar standard deviation of Dice score in common CHD and less common CHD compared with Seg-CNN \cite{payer2017multi}.

\begin{figure*}[!tb]
\includegraphics[width=\textwidth]{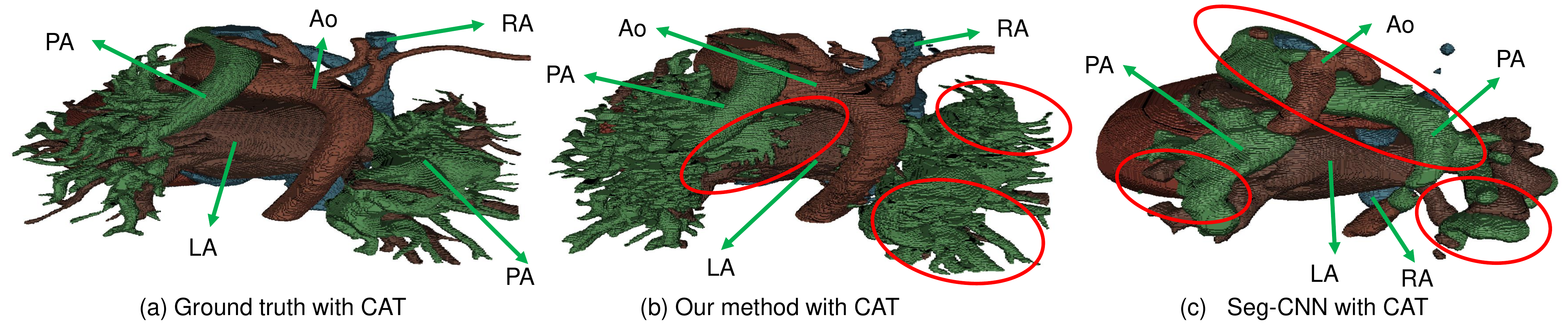}
\vspace{-20pt}
\caption{Visualized comparison between the state-of-the-art method Seg-CNN \cite{payer2017multi} and our method. The differences from the ground truth are highlighted by the red circles. 
} 
\vspace{-2pt}
\label{fig_compare}
\end{figure*}

Visualization of CAT segmentation using our method and Seg-CNN is shown in Fig. \ref{fig_compare}.
Our method can clearly segment Ao and PA with some slight mis-segmentation between PA and LA.
However, Seg-CNN segments the main part of Ao as PA, which is due to the fact that pixel-level segmentation by U-net is only based on the surrounding pixels, and the connection information is not well exploited. 

\begin{figure*}[!tb]
\centering
\includegraphics[width=0.9\textwidth]{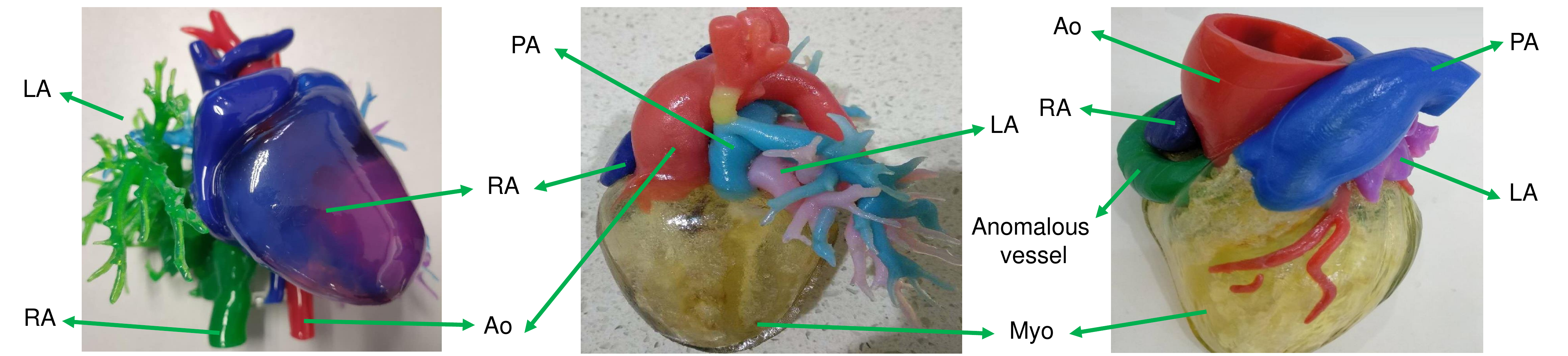}
\vspace{-6pt}
\caption{Examples of 3D printing models using our method with some minor manual refinement. 
} 
\vspace{-12pt}
\label{fig_3d_printing}
\end{figure*}

{Examples of 3D printing models using our method with some minor manual refinement are shown in Fig. \ref{fig_3d_printing}.
We can notice that the printed model is with correct and clear shape and connections, and experienced radiologists have confirmed its usability to clinic use.
Note that the refinement is mainly about adding some thin but critical vessels such as coronary vessels.}

\section{Conclusion}
In this paper we proposed a whole heart and great vessel segmentation framework for 3D printing of CT images in CHD. 
We first used deep learning to segment the four chambers and myocardium followed by blood pool, where variations are usually small. We then extracted the connection information and apply graph matching to determine the categories of all the vessels.
We collected a CHD dataset in CT with 68 3D images, and the ground truth has seven categories: LV, RV, LA, RA, myocardium, Ao and PA. Totally 14 types of CHD are included in this dataset which is made publicly available.
Compared with the state-of-the-art method for whole heart and great vessel segmentation in normal anatomy, our method can achieve 11.9\% improvement in Dice score on average.
{We also printed out part of the segmentation results with minor manual refinement, and showed that it can be applied to clinic use. }

\section*{Acknowledgment}

This work was supported by the National Key Research and Development Program of China [2018YFC1002600], Science and Technology Planning Project of Guangdong Province, China [No.2017A070701013 and 2017B090904034, 2017030314109, 2019B020230003], and Guangdong peak project [DFJH201802].


\begin{thebibliography}{10}
\providecommand{\url}[1]{#1}
\csname url@samestyle\endcsname
\providecommand{\newblock}{\relax}
\providecommand{\bibinfo}[2]{#2}
\providecommand{\BIBentrySTDinterwordspacing}{\spaceskip=0pt\relax}
\providecommand{\BIBentryALTinterwordstretchfactor}{4}
\providecommand{\BIBentryALTinterwordspacing}{\spaceskip=\fontdimen2\font plus
\BIBentryALTinterwordstretchfactor\fontdimen3\font minus
  \fontdimen4\font\relax}
\providecommand{\BIBforeignlanguage}[2]{{%
\expandafter\ifx\csname l@#1\endcsname\relax
\typeout{** WARNING: IEEEtran.bst: No hyphenation pattern has been}%
\typeout{** loaded for the language `#1'. Using the pattern for}%
\typeout{** the default language instead.}%
\else
\language=\csname l@#1\endcsname
\fi
#2}}
\providecommand{\BIBdecl}{\relax}
\BIBdecl

\bibitem{bhat2016illustrated}
V.~Bhat, V.~BeLaVaL, K.~Gadabanahalli, V.~Raj, and S.~Shah, ``Illustrated
  imaging essay on congenital heart diseases: multimodality approach part i:
  clinical perspective, anatomy and imaging techniques,'' \emph{Journal of
  clinical and diagnostic research: JCDR}, vol.~10, no.~5, p. TE01, 2016.

\bibitem{xu2018resource}
X.~Xu, T.~Wang, Q.~Lu, and Y.~Shi, ``Resource constrained cellular neural
  networks for real-time obstacle detection using fpgas,'' in \emph{2018 19th
  International Symposium on Quality Electronic Design (ISQED)}.\hskip 1em plus
  0.5em minus 0.4em\relax IEEE, 2018, pp. 437--440.

\bibitem{xu2018quantization}
X.~Xu, Q.~Lu, L.~Yang, S.~Hu, D.~Chen, Y.~Hu, and Y.~Shi, ``Quantization of
  fully convolutional networks for accurate biomedical image segmentation,'' in
  \emph{Proceedings of the IEEE Conference on Computer Vision and Pattern
  Recognition}, 2018, pp. 8300--8308.

\bibitem{xu2018scaling}
X.~Xu, Y.~Ding, S.~X. Hu, M.~Niemier, J.~Cong, Y.~Hu, and Y.~Shi, ``Scaling for
  edge inference of deep neural networks,'' \emph{Nature Electronics}, vol.~1,
  no.~4, p. 216, 2018.

\bibitem{tianchen2019}
T.~Wang, J.~Xiong, X.~Xu, and Y.~Shi, ``Scnn: A general distribution based
  statistical convolutional neural networkwith application to video object
  detection,'' in \emph{AAAI’19}, 2019.

\bibitem{xu2018efficientHardware}
X.~Xu, Q.~Lu, T.~Wang, Y.~Hu, C.~Zhuo, J.~Liu, and Y.~Shi, ``Efficient hardware
  implementation of cellular neural networks with incremental quantization and
  early exit,'' \emph{JETC}, vol.~14, no.~4, p.~48, 2018.

\bibitem{xu2017edge}
X.~Xu, Q.~Lu, T.~Wang, J.~Liu, C.~Zhuo, X.~S. Hu, and Y.~Shi, ``Edge
  segmentation: Empowering mobile telemedicine with compressed cellular neural
  networks,'' in \emph{ICCAD}.\hskip 1em plus 0.5em minus 0.4em\relax IEEE
  Press, 2017, pp. 880--887.

\bibitem{li2019exploiting}
B.~Li, C.~Chenli, X.~Xu, T.~Jung, and Y.~Shi, ``Exploiting computation power of
  blockchain for biomedical image segmentation,'' \emph{arXiv preprint
  arXiv:1904.07349}, 2019.

\bibitem{liu2019compression}
Z.~Liu, X.~Xu, T.~Liu, Q.~Liu, Y.~Wang, Y.~Shi, W.~Wen, M.~Huang, H.~Yuan, and
  J.~Zhuang, ``Machine vision guided 3d medical image compression for efficient
  transmission and accurate segmentation in the clouds,'' in \emph{CVPR}, 2019,
  pp. 8300--8308.

\bibitem{zhuang2016multi}
X.~Zhuang and J.~Shen, ``Multi-scale patch and multi-modality atlases for whole
  heart segmentation of mri,'' \emph{Medical image analysis}, vol.~31, pp.
  77--87, 2016.

\bibitem{wang2018two}
C.~Wang, T.~MacGillivray, G.~Macnaught, G.~Yang, and D.~Newby, ``A two-stage 3d
  unet framework for multi-class segmentation on full resolution image,''
  \emph{arXiv preprint arXiv:1804.04341}, 2018.

\bibitem{payer2017multi}
C.~Payer, D.~{\v{S}}tern, H.~Bischof, and M.~Urschler, ``Multi-label whole
  heart segmentation using cnns and anatomical label configurations,'' in
  \emph{International Workshop on Statistical Atlases and Computational Models
  of the Heart}.\hskip 1em plus 0.5em minus 0.4em\relax Springer, 2017, pp.
  190--198.

\bibitem{xu2018cfun}
Z.~Xu, Z.~Wu, and J.~Feng, ``Cfun: Combining faster r-cnn and u-net network for
  efficient whole heart segmentation,'' \emph{arXiv preprint arXiv:1812.04914},
  2018.

\bibitem{cciccek20163d}
{\"O}.~{\c{C}}i{\c{c}}ek, A.~Abdulkadir, S.~S. Lienkamp, T.~Brox, and
  O.~Ronneberger, ``3d u-net: learning dense volumetric segmentation from
  sparse annotation,'' in \emph{International conference on medical image
  computing and computer-assisted intervention}.\hskip 1em plus 0.5em minus
  0.4em\relax Springer, 2016, pp. 424--432.

\bibitem{yu20163d}
L.~Yu, X.~Yang, J.~Qin, and P.-A. Heng, ``3d fractalnet: dense volumetric
  segmentation for cardiovascular mri volumes,'' in \emph{Reconstruction,
  segmentation, and analysis of medical images}.\hskip 1em plus 0.5em minus
  0.4em\relax Springer, 2016, pp. 103--110.

\bibitem{wolterink2016dilated}
J.~M. Wolterink, T.~Leiner, M.~A. Viergever, and I.~I{\v{s}}gum, ``Dilated
  convolutional neural networks for cardiovascular mr segmentation in
  congenital heart disease,'' in \emph{Reconstruction, segmentation, and
  analysis of medical images}.\hskip 1em plus 0.5em minus 0.4em\relax Springer,
  2016, pp. 95--102.

\bibitem{pace2018iterative}
D.~F. Pace, A.~V. Dalca, T.~Brosch, T.~Geva, A.~J. Powell, J.~Weese, M.~H.
  Moghari, and P.~Golland, ``Iterative segmentation from limited training data:
  Applications to congenital heart disease,'' in \emph{Deep Learning in Medical
  Image Analysis and Multimodal Learning for Clinical Decision Support}.\hskip
  1em plus 0.5em minus 0.4em\relax Springer, 2018, pp. 334--342.

\bibitem{lajevardi2013retina}
S.~M. Lajevardi, A.~Arakala, S.~A. Davis, and K.~J. Horadam, ``Retina
  verification system based on biometric graph matching,'' \emph{IEEE
  transactions on image processing}, vol.~22, no.~9, pp. 3625--3635, 2013.

\end{thebibliography}
\end{document}